 \documentclass[oldversion]{aa}			

\usepackage{graphicx}
\usepackage{txfonts}

\begin{document}

\title{The optical to $\gamma$-ray emission of the Crab pulsar: \\
a multicomponent model}

\author{E. Massaro\inst{1} 
\and R. Campana\inst{1} 
\and G. Cusumano\inst{2} 
\and T. Mineo\inst{2} }
	
\institute{Dipartimento di Fisica, Universit\`a di Roma ``La Sapienza'', Piazzale 
A. Moro 2, I-00185, Roma, Italy 
\and
INAF-IASF, Sezione di Palermo, Via Ugo La Malfa 153, I-90146, Palermo,
Italy }

\offprints{~~~~~~~~\\
E.~Massaro: ~enrico.massaro@uniroma1.it}

\date{Received:.; accepted:.}

\titlerunning{A multicomponent model of the Crab pulsar}
\authorrunning{E. Massaro et al.}

\abstract
{We present a multicomponent model to explain the 
features of the pulsed emission and spectrum of the Crab Pulsar, on the basis of  
X and $\gamma$-ray observations obtained with BeppoSAX, INTEGRAL and CGRO.
This model explains the evolution of the pulse shape and of the 
phase-resolved spectra, ranging from the optical/UV to the GeV energy band, on 
the assumption that the observed emission is due to more components.
The first component, $C_O$, is assumed to have the pulsed double-peaked 
profile observed at the optical frequencies, while the second component, 
$C_X$, is dominant in the interpeak and second peak phase regions. 
The spectra of these components are modelled with log-parabolic laws and 
their spectral energy distributions have peak energies at 12.2 and 178 keV, 
respectively.
To explain the properties of the pulsed emission in the MeV-GeV band, we 
introduce two more components, $C_{O\gamma}$  and $C_{X\gamma}$, with phase 
distributions similar to those of $C_O$ and $C_X$ and log-parabolic spectra 
with the same curvature but peak energies at about 300 MeV and 2 GeV. 
This multicomponent model is able to reproduce both the broadband phase-resolved 
spectral behaviour and the changes of the pulse shape with energy. 
We also propose some possible physical interpretations in which $C_O$ and $C_X$ 
are emitted by secondary pairs via a synchrotron mechanism while $C_{O\gamma}$ and
$C_{X\gamma}$ can originate either from Compton scattered or primary curvature
photons.}


\maketitle

\section{Introduction}
The origin of the high energy emission of rotation-powered pulsars is
still an unsolved problem. One of the main difficulties is related to
the description of the phase and energy distributions of the pulsed
signal, which depends on both a physical and geometrical modelling
of the magnetosphere.  
It is still unclear, for example, whether electrons (or positrons) are 
accelerated and radiate streaming out the polar cap regions (Ruderman 
\& Sutherland 1975, Salvati \& Massaro 1978, Sturner \& Dermer 1994, 
Daugherty \& Harding 1994, 1996; Muslimov \& Harding 2003) or in the 
outer gaps (Cheng, Ho \& Ruderman 1986a,b; Chiang \& Romani 1994, 
Romani \& Yadigaroglu 1995, Cheng, Ruderman \& Zhang 2000; 
Zhang \& Cheng 2002, Hirotani, Harding \& Shibata 2003). 
Another important problem concerns how quantum processes, 
like magnetic pair production (\cite{erber66}) and photon splitting 
(\cite{adler71}), modify the high energy $\gamma$-ray spectrum.

The Crab pulsar (PSR B0531+21) has been the best studied object of this class since 
its discovery (Staelin \&  Reifenstein 1968) and the amount of data collected 
is rich enough to search for a detailed physical picture of its 
emission properties.
It is well known that the pulse shape of Crab has a characteristic 
double peak structure, with a phase separation of 0.4, detected from 
the radio band to $\gamma$ rays and changing with energy.
A very remarkable feature is that the so-called second peak, hereafter P2, 
in the X and soft $\gamma$-ray ranges becomes progressively higher than the 
first peak (P1). A similar increase is also evident in the emission
between the peaks, usually named the Interpeak region (Ip) or bridge
(see Fig. 1).
Above about 10 MeV, P1 is again the dominant feature.
A satisfactory explanation for these changes has not been found so far. 

On the basis of high quality BeppoSAX data, covering a wide energy 
range from 0.1 to about 300 keV, we proposed a two component model (Massaro 
et al. 2000, hereafter MCLM) to interpret this behaviour. 
In the same paper we studied in detail the energy spectrum of the core of P1
(corresponding to a phase interval having a width of only 0.027 around the 
maximum) which shows a continuous steepening at high energies.
We found that in the energy range 0.1--300 keV this spectral distribution is 
well represented by a parabolic law in a double logarithmic plot (hereafter 
log-parabola) with a rather mild curvature. 
The extrapolation of this model in the $\gamma$-ray range, however, fails 
to reproduce the data and a more complex modelling is required.

\begin{figure*}
\centering
\resizebox{\hsize}{!}{\includegraphics[angle=90]{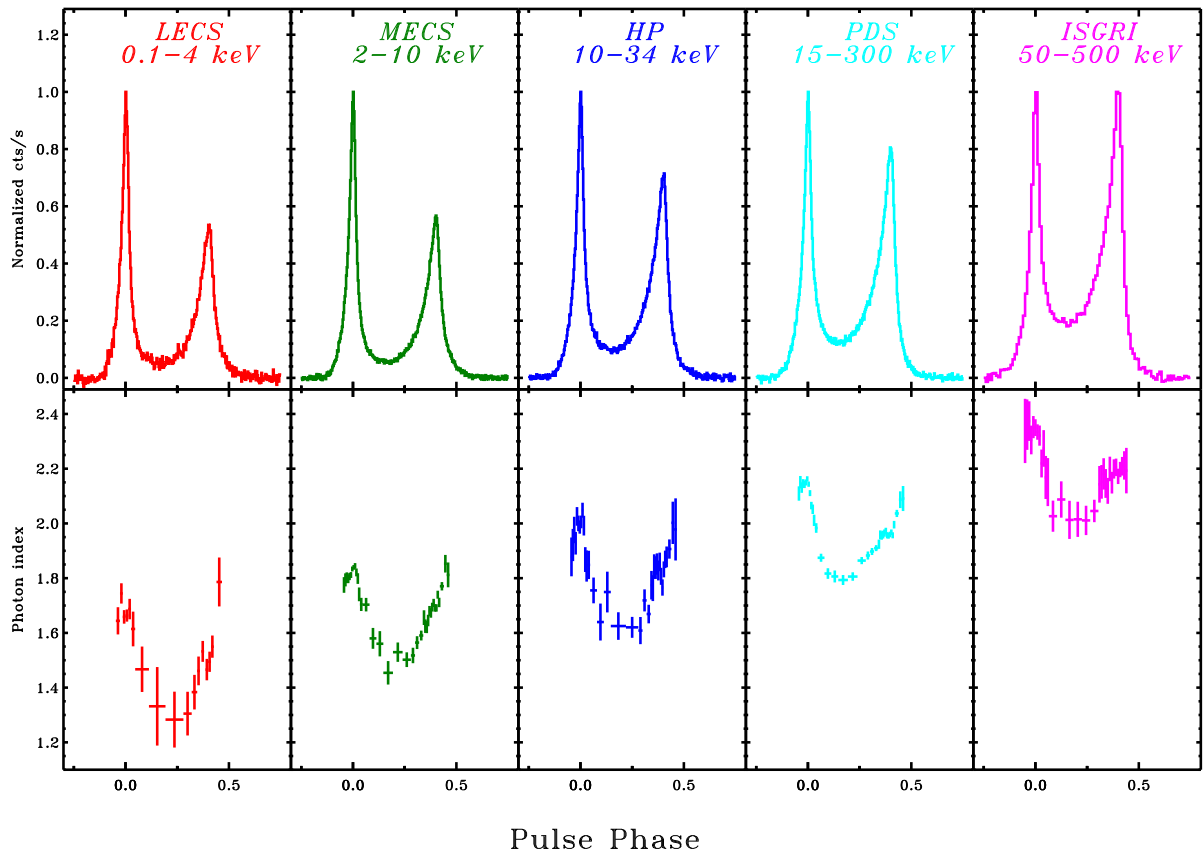}}
\caption{The pulse profile and the phase dependent photon index of 
the Crab pulsar observed with the four Narrow Field Instruments of 
BeppoSAX and with ISGRI-INTEGRAL experiment. 
The respective energy ranges are indicated in the upper 
panels. 
All the profiles are normalized to unity at the maximum of P1.
Note the change of relative intensity of P2 and Ip with respect to P1
and the increase of the photon indices.
}
\label{fig1}
\end{figure*}

In a subsequent paper Kuiper et al. (2001) introduced three components to
describe the spectrum up to the $\gamma$-ray data obtained with COMPTEL 
and EGRET on board ComptonGRO.
These authors based the analysis mainly on finding best fits of the
spectral distributions in rather narrow phase intervals and found
that the spectral variation of the pulsed emission with phase can
be modelled by two log-parabolic components 
with a relative normalization changing with phase.
A further third power law component, having a photon index equal
to 2.07, was necessary for the emission of P1 and P2 to
to match the EGRET data at energies higher than about 10 MeV. \\
In this paper we develop a model able to describe the phase and spectral 
distributions of the emission over a frequency interval from the optical
frequencies to GeV range. This model is an extension of that presented in
MCLM and it based on the results of a new detailed data analysis of 
many BeppoSAX observations which includes PDS data from March 1999 to
April 2001 not considered by MCLM. 
The timing accuracy has been verified using detailed pulse profiles 
obtained from RXTE archive data.
Moreover, to extend the energy range, we analysed several more recent 
observations performed with the IBIS-ISGRI experiment on board the 
INTEGRAL satellite and considered the results of Kuiper et al. (2001) 
on the COMPTEL and EGRET observations up to a few GeV. 
The main goal of our work is the definiton of a scenario that
can be used to develop more detailed physical models
of the Crab pulsar high-energy emission. 

\section{Observations and data reduction}

\subsection{BeppoSAX}
BeppoSAX observed the Crab Nebula and its pulsar on several occasions 
during its lifetime, because this very bright source was used
to monitor the responses of the four Narrow Field Instruments (NFI).
All the data collected up to 1998 were used in MCLM 
and a complete description of their reduction is presented
in that work.
In the new analysis presented in this paper we did not consider 
some early PDS observations, particularly those performed on 31 August 1996, 
during the Science Verification Phase, and on 6 April 1998 because the
detector gain was found to be not precisely set,
whereas additional PDS data were obtained from other observations performed 
between 1999 and 2001, corresponding to an increase of the exposure time 
of about 12\% with respect to MCLM.
The complete log of all BeppoSAX pointings used in the present work is given 
in Table 1.

For the imaging instruments we selected all the events within circular
regions centered on the source position and having radii of 4\arcmin\ 
(MECS) and 8\arcmin\ (LECS). 
This choice corresponds to using about 90\% of the total 
source signal in both instruments, but it allows us to apply the best 
tested spectral response matrices. 

Phase histograms of the Crab pulsar were evaluated for each NFI and
each pointing using the period folding technique. 
The UTC arrival times of all selected events were converted to the
Solar System Barycentre with the DE200 ephemeris.
The values of $P$ and $\dot P$ for each observation epoch were derived
from the Jodrell Bank Crab Pulsar Monthly 
Ephemeris\footnote{\texttt{http://www.jb.man.ac.uk/}.}. 
We constructed a large set of 300 bin phase histograms for each 
energy channel of each NFI.
The zero phase was fixed at the centre of the first peak, evaluated by Gaussian fits. 
All the histograms for the same energy channel of each NFI were 
then added.
Before this operation we verified that all the profiles of the various
observation epochs had fully compatible shapes and therefore similar  
folding accuracy.
A summary of all these phase histograms is shown in the upper panels of
Fig.~1.

A relevant point we stress when working with spectra
obtained with several instruments is that a proper evaluation of the
inter-calibration factors is required.
For the BeppoSAX NFIs, the accurate ground
and in-flight calibrations were used to establish the admissible 
ranges for the factors between the MECS and the other
three instruments: $k_{ML}$ for the LECS, $k_{MH}$ for the HPGSPC and 
$k_{MP}$ for the PDS (Fiore et al. 1999).
When performing the spectral analysis, we left these three parameters free
and found the values $k_{ML}$=0.74 and $k_{MH}$=0.88,
within the ranges given by Fiore et al. (1999).   
For point sources these authors gave 0.77 $\leq k_{MP}\leq$0.93, reduced 
to 0.86$\pm$0.03 for sources with a PDS count rate higher than 2 ct/s.
We found in our best fits $k_{MP}$=0.82, very close to the above prescription, 
and that we considered fully satisfactory because it was derived with 
the log-parabolic model instead of the simple power law used by Fiore et al. 
(1999).
 
\begin{table}
\caption{Log of BeppoSAX NFIs Crab Pulsar pointing epochs and net exposure times.}
\label{tab:bsax}
\begin{flushleft}
\begin{tabular}{crrrr}
\hline
\multicolumn{1}{c}{Observation } & \multicolumn{4}{c}{Exposure Times (s)} \\
\multicolumn{1}{c}{Date} & \multicolumn{1}{c}{LECS} & \multicolumn{1}{c}{MECS$^a$} &
 \multicolumn{1}{c}{HPGSPC}  & 
 \multicolumn{1}{c}{PDS} \\ 
\hline
31~Aug~1996     & --      &  --     &  28\,384 &  -- \\ 
~6~Sep~1996     & 5\,733  & 33\,482 & --       &  --   \\
30~Sep~1996     & --      &  7\,874 &  --      & 3596  \\ 
12~Apr~1997     & 1\,257  & 18\,551 &  9\,344  &  9\,338 \\
~8~Oct~1997 & 12\,096 & 30\,769 &  9\,158  & 19\,673 \\
~6~Apr~1998 & 8\,730  & 28\,694 &  12\,732 & -- \\ 
13~Oct~1998     & --      &  --     &   --     & 17\,476 \\ 
~9~Mar~1999     & --      &  --     &   --     & 13\,674  \\ 
25~Sep~1999     & --      &  --     &   --     & 15\,417 \\
10~Apr~2000     & --      &  --     &   --     & 15\,602  \\
~4~Apr~2001     & --      &  --     &   --     & 7\,597  \\
\hline
Total exposure  & 27\,816 & 119\,370 & 59\,618 & 102\,373  \\ 
\hline 
\\
\end{tabular}
$^a$~MECS operated with two detectors after May 1997.\\
\end{flushleft}
\end{table}

\begin{table}
\caption{Log of INTEGRAL-ISGRI observations of the Crab and of net exposure times.}
\label{tab:isgri}
\begin{center}
\begin{tabular}{cr}
\hline
Start -- Stop day & Exposure Times (s)~~~~~\\
\hline
 07~Feb~2003 -- 09~Feb~2003  &  142\,300 \\
 10~Feb~2003 -- 11~Feb~2003  &  56\,000  \\
 16~Feb~2003 -- 18~Feb~2003  &  61\,600  \\
 19~Feb~2003 -- 21~Feb~2003  &  30\,700  \\
 22~Feb~2003 -- 24~Feb~2003  &  9\,800  \\
 25~Feb~2003 -- 26~Feb~2003  &  2\,100 \\
 15~Aug~2003 -- 17~Aug~2003  &  37\,900 \\
 17~Aug~2003 -- 17~Aug~2003  &  21\,600 \\
\hline
Total exposure & 362\,000 \\
\hline
\end{tabular}
\end{center}
\end{table}

\subsection{RossiXTE}
To verify the accuracy of BeppoSAX timing, we compared the pulse
profiles as observed by MECS with that obtained by the PCA instrument 
(Jahoda et al. 1996) on board of the Rossi X-Ray Timing Explorer.
The latter instrument has a timing accuracy of 2 $\mu$s 
(Rots et al. 2004) with respect to the spacecraft clock and an absolute 
time accuracy of 5-8 $\mu$s with respect to UTC.
We used the RXTE observation of Crab performed on December 3, 2004.
PCA data were obtained in an "event mode" that provided 128 energy
channels with 250 $\mu$s time resolution.
PCA data were screened following standard selection criteria excluding
time intervals corresponding to South Atlantic Anomaly passages, Earth
limb closer than 10 degrees and with angular distance between the source
and pointing direction larger than 0$^{\circ}$.02. 
The total exposure time after all screening criteria was 709 s.
After the conversion of the arrival times to the Solar System Barycentre
by using the DE200 ephemeris, phase histograms were obtained by
folding data with the Jodrell Bank Crab Pulsar Monthly Ephemeris as before.

Fig. 2 shows the RossiXTE-PCA pulse profile, extracted in the 2.6-4.1
keV energy band and with 300 phase bins (110 $\mu$s), together with that 
obtained in the same energy range with MECS.
Both profiles have been normalized to unity at the maximum of P1, after 
subtraction of the off-pulse constant level (0.6--0.83) and arbitrarily 
shifted to set the zero phase at the centre of the first peak.
The comparison shows that the peak widths between the two
instruments are very similar. 
The two profiles have differences of only a few percent in amplitude 
(Fig. 2, lower panel) and there is no evidence of a systematic effect on 
the timing in the MECS data.
We conclude that the BeppoSAX profiles accumulated over multiple
observations do not undergo any significant broadening
due either to the clock time assignment or phase misalignment among the
different histograms added together.

\begin{figure}
\resizebox{\hsize}{!}
{\includegraphics[height=8cm,width=8cm,angle=0]{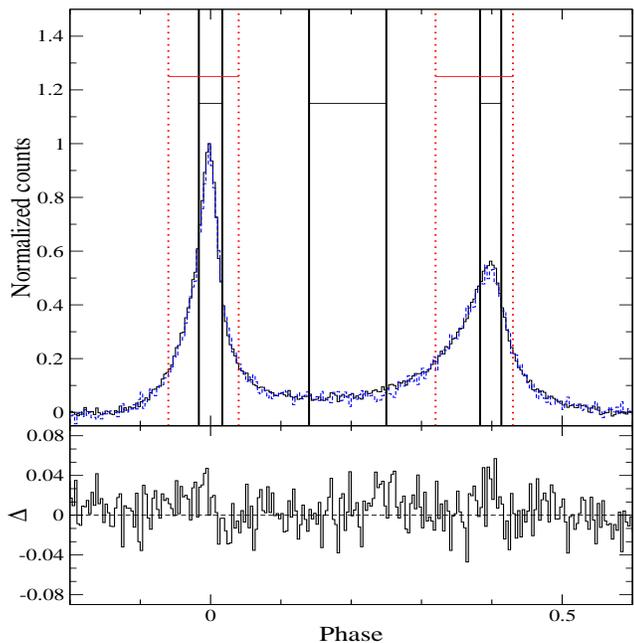}}
\caption{The comparison of the normalized pulse profiles in the 2.6--4.1 keV 
energy range of BeppoSAX-MECS (solid histogram) and RossiXTE-PCA (dashed 
histogram) (upper panel). 
The difference plot in the lower panel has an amplitude
of only a few percent and no sign of systematic effects on the timing. 
In the upper panel we also show the phase intervals of P1, Ip and P2 used 
in our spectral analysis (solid vertical lines - black) and according 
to Kuiper et al. (2001) (dotted vertical lines - red) used to compute
the broad-band spectral distributions and the P2/P1 and Ip/P1 ratios
(see Sect. 4).
}
\label{fig2}
\end{figure}

\subsection{INTEGRAL}

To gain more information at energies higher than 200 keV we considered 
also the recent data obtained with IBIS-ISGRI (Lebrun et al. 2003) on 
board the INTEGRAL satellite.
In our analysis we considered only the IBIS observations having an off-axis 
angle not larger than 1\degr\ which are listed in Table \ref{tab:isgri}.
Phase histograms with 100 bins were obtained for each 
pointing using, as before, the period folding technique (see Fig. 1).
The accuracy of INTEGRAL timing with respect to RXTE has been
verified by Kuiper et al. (2003) 
The zero phase was shifted at the centre of the first peak, coherently with 
the assumption for the other instruments.
A more complete spectral analysis of these data, together with those obtained 
using other INTEGRAL instruments, can be found in Mineo et al. (2006).

\section{The interstellar absorption in the Crab direction}

\subsection{X-ray absorption}
An important piece of information for the study of the intrinsic shape of the 
soft X-ray spectral distribution is the value of the equivalent hydrogen 
column density and the composition of the interstellar matter in the 
Crab direction.
In MCLM we assumed $N_{\rm H}$=3.2$\times$10$^{21}$ cm$^{-2}$ and a 
mean solar composition. 
This value is higher than those considered in previous papers, usually
lower than 3$\times$10$^{21}$ cm$^{-2}$ (see, for instance, Toor \& 
Seward 1977, Ride \& Walker 1977, Pravdo \& Serlemitsos 1981) and 
in agreement with the analysis of ASCA data by Fukuzawa et al. (1997) and 
with the result of Sollerman et al. (2000) who estimated 
$N_{\rm H}=3.2\pm0.5\times10^{21}$ cm$^{-2}$ from STIS/HST observations 
of the Ly$\alpha$ line.
This column density, however, was not confirmed by subsequent studies 
that indicated an even larger $N_{\rm H}$ and different chemical abundances. 
Willingale et al. (2001) on the basis of EPIC-MOS data found a column
density of 3.45$\times$10$^{21}$ cm$^{-2}$ with an underabundance of 
oxygen and iron of 63\% with respect to the solar composition. 
Weisskopf et al. (2004) obtained $N_{\rm H}=4.2\times$10$^{21}$ cm$^{-2}$
with the abundances given by Wilms et al. (2000) and using a photoelectric 
absorption model that includes the scattering due to interstellar grains.

\begin{figure}
\resizebox{\hsize}{!}{\includegraphics[angle=-90]
{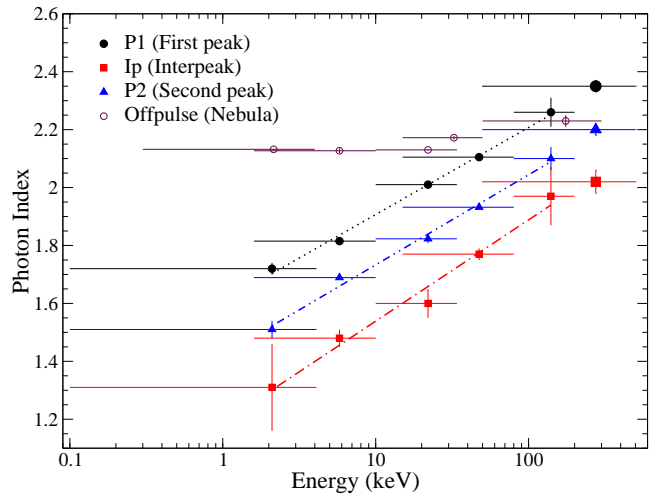}}
\caption{The energy dependence of the photon indices of P1, Ip and P2, 
evaluated in the ranges of the LECS, MECS , HPGSPC, two PDS bands
(15--80, 80--200 keV) and INTEGRAL-ISGRI (50--500 keV, large symbols).
The off-pulse photon index, representative of the nebular emission, is also 
shown. Best-fit linear interpolations of BeppoSAX photon indices of 
P1, Ip and P2 are plotted.
}
\label{fig3}
\end{figure}

In order to use a more precise estimate of $N_{\rm H}$ and the most 
suited chemical composition we performed a new analysis of the X-ray 
spectrum of the DC emission of Crab (pulse phases 0.60--0.83) using 
LECS and MECS data combined.
Background, estimated from archive blank fields, was subtracted from the
events.
We adopted a simple power law ($F(E)=K(E/E_0)^{-\Gamma}$) as the best fitting 
model because in this band there is no evidence of a change of the 
nebular spectral index with energy (see next Section). 
The considered abundances and photoelectric cross sections were those 
of the models \texttt{aneb+wabs} (Anders \& Ebihara 1982, Morrison 
\& McCammon 1983) and \texttt{angr+bcmc} (Anders \& Grevesse 1989, 
Balucinska-Church \& McCammon 1992, \texttt{vphabs}) all implemented 
in the XSPEC package. 
The results are reported in Table \ref{tab:nh}, where for each model 
we give the best fit values of $N_{\rm H}$ and of the photon index 
$\Gamma$, the fraction of oxygen and iron with respect to the default 
value and the reduced $\chi^2$. 
All the fits were acceptable (reduced $\chi^2 < 1$) and the column 
densities were systematically higher than 3.2$\times$10$^{21}$ 
cm$^{-2}$. 
Photon index values were found to be very stable with a mean value of 2.124. 
We also found (see Sect. 6.2) that the residuals at energies lower than 
0.8 keV had a systematic deviation which disappeared using column density
values around 3.6$\times$10$^{21}$ cm$^{-2}$. 
We decided, therefore, to adopt in our analysis this $N_{\rm H}$ value 
together with the \texttt{angr+bcmc} model, as indicated by other instruments 
with higher spectral capabilities.

\subsection{Optical-UV extinction}
The high estimate of the equivalent column density and the possible 
difference of the chemical composition with respect to the solar one could also affect the
optical-UV extinction in the direction of Crab.
The analysis of Sollerman et al. (2000)  converges around a value of $E(B-V) = 0.50$ mag 
and indicates that the extinction curve closely follows the standard one.
Thus the $N_{\rm H}/E(B-V)$ ratio is equal to 6.9$\times$10$^{21}$ 
cm$^{-2}$ mag$^{-1}$, about 30\% higher than the value reported by Predehl \&
Schmitt (1995). 

Such a relevant optical-UV intrinsic extinction makes the evaluation
of the pulsed flux and spectrum quite difficult at these frequencies.
Several authors (e.g. Percival et al. 1993, Sollerman et al. 2000) agree that 
the dereddened spectrum is a power law with a spectral index close to zero.
The difference to uncorrected data is quite large: in fact,
the observed spectral index is around $-2$, and small differences in the
values of $E(B-V)$ or $A_V$ can modify the dereddened slope and flux.
Of course, the extrapolation of UV data in the X-ray range may be affected by large
uncertainties. 

\begin{table*}
\caption{Estimates of the equivalent hydrogen column density $N_{\rm H}$, 
from combined LECS (0.1--4 keV) and MECS (1.6--10 keV) observations with 
various absorption models and chemical abundances.
}
\label{tab:nh}
\begin{center}
\begin{tabular}{cccccc}
\hline
\multicolumn{1}{c}{Model} & \multicolumn{1}{c}{$N_{\rm H}(^*)$} & \multicolumn{1}{c}{$\Gamma$} & \multicolumn{1}{c}{$F_O$} & \multicolumn{1}{c}{$F_{Fe}$} & \multicolumn{1}{c}{$\chi^2_r$}\\
\hline
\texttt{aneb+wabs} & 3.54 $\pm$ 0.01 & 2.124 $\pm$ 0.001 & --            & --            & 0.89 \\
\texttt{angr+bcmc}  & 3.33 $\pm$ 0.01 & 2.124 $\pm$ 0.001 & --            & --            & 0.99 \\
\texttt{angr+bcmc}  & 3.62 $\pm$ 0.04 & 2.125 $\pm$ 0.001 & 0.81 $\pm$ 0.02 & --            & 0.90 \\
\texttt{angr+bcmc}  & 3.63 $\pm$ 0.04 & 2.124 $\pm$ 0.001 & 0.83 $\pm$ 0.03 & 0.87 $\pm$ 0.05 & 0.88 \\
\texttt{angr+bcmc}  & 3.42 $\pm$ 0.02 & 2.123 $\pm$ 0.001 & --            & 0.74 $\pm$ 0.04 & 0.93 \\
\hline
\\
$(^*)$ In units of 10$^{21}$ cm$^{-2}$.
\end{tabular}
\end{center}
\end{table*}

\section{Phase resolved X-ray spectroscopy}

In MCLM we showed that a simple power law does not provide an acceptable 
representation of the spectral distribution of the central bins of P1, 
corresponding to the narrow phase interval 0.99--0.01667. 
This is not a specific property of P1 but can be extended to the entire
pulsed signal as apparent from the phase resolved spectroscopy of Fig. 
\ref{fig1},
where pulse profiles and best fit photon indices evaluated in the energy 
bands of the four NFIs of BeppoSAX and of ISGRI-INTEGRAL are shown. 
Upper panels show the pulse profiles while lower panels show the phase 
evolution of the photon indices: note the well known increase of Ip and P2 
with respect to P1 and that photon indices are also increasing with 
energy: that of P1 changes from about 1.6 to 2.4 and that of the Ip 
from 1.3 to 1.95.
The power law fit was applied to the events after the subtraction of a
DC level, estimated from the phase interval 0.60--0.83, and assumed 
to be representative of the nebular continuum.
High resolution X-ray images obtained with Chandra (Tennant et al. 
2001) showed that there is a weak emission from the pulsar also in this 
phase interval. 
Its mean intensity is of the order of few percent than that 
of the peaks and becomes negligible compared to the nebular emission
when considering data extracted from wider angular beams, as in our case.  

\begin{figure}
\resizebox{\hsize}{!}{\includegraphics[height=8cm,width=8cm,angle=-90]
{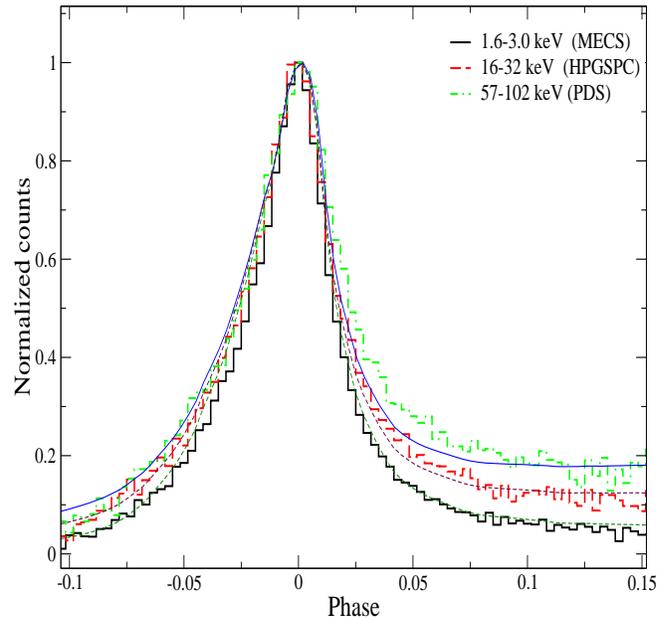}}
\caption{The shape of P1 in three energy ranges. 
Maxima have been normalized to unity for a better comparison. 
Note the broadening of the peak with a marked asymmetry on the right side.
Solid lines represent the P1 profiles derived using the two component 
model (see Sect. 6.3).
}
\label{fig4}
\end{figure}

To study the behaviour of the photon index with energy in detail we 
considered the following three phase intervals representative
of P1 (0.9833--0.0167), the Interpeak (Ip) region (0.14--0.25) and 
P2 (0.3833--0.4137). 
This choice was made to have intervals wide enough to obtain estimates
of $\Gamma$ with small statistical uncertainties and sufficiently narrow
to consider the emission of P1 mainly due to one of the components
introduced by MCLM.
In subsequent analysis, when comparing the results of our multicomponent
model with literature data (see Sect. 7), it was necessary to consider 
for P1 and P2 phase intervals broader than those given above and 
we adopted the same definition given by Kuiper et al. (2001).
The phase interval of Ip was the same.
Fig. 2 shows the differences between these two sets.
 
The resulting values of $\Gamma$ in the narrow phase intervals are shown 
in Fig. \ref{fig3}, together with those of the off-pulse, assumed originating
in the nebula.  
The values of the P1 photon index vary from 1.72$\pm$0.02 in the
0.1--4 keV range to 2.30$\pm$0.05 at (80--200 keV); 
correspondingly, those of Ip vary from 1.31$\pm$0.10 to 1.98$\pm$0.10 and 
those of P2 from 1.51$\pm$0.03 to 2.14$\pm$0.05.
We considered also INTEGRAL-ISGRI data in the rather broad interval 50--500 
keV and obtained photon indices fully consistent with the increasing
trend of BeppoSAX values.
The nebular spectrum has a remarkably constant value equal to 2.131$\pm$0.002, 
up to the PDS range where it slightly increases to 2.172$\pm$0.006 (15--80 keV)
and to 2.23$\pm$0.02 (80--200 keV). Note that the last value agrees very
well with that obtained by Kuiper et al. (2001) from COMPTEL data between
0.75 and 30 MeV.
The very good consistency of the nebular photon index measured in LECS,
MECS and HPGSPC is an indication of the high accuracy of their calibration
and of the correct choice of the intercalibration constants.

It is clear from these results that a law more complex than a simple
power law must be used to describe the broad-band spectral properties 
of the Crab pulsar when the data of all the NFIs are considered.
A simple power law must be rejcted because of the high values 
of the reduced $\chi^2$, found always larger than 2.

In the MCLM analysis of the P1 spectrum, good spectral fits are obtained 
using a second order law in the double log representation:
\begin{equation}\label{eq1}
F(E)=A~(E/E_0)^{-(a+b~\mathrm{Log}(E/E_0))} \, \, ,
\end{equation}
where $E_0$ is taken equal to 1 keV, and therefore $a$ corresponds to
the photon index at this energy, while $b$ measures the curvature of
the spectral distribution.
We can also define an energy dependent photon index: 
\begin{equation}\label{eq2}
\Gamma(E) = -\frac{d \mathrm{Log}F(E)}{d \mathrm{Log}E} = a + 2 b~\mathrm{Log}(E/E_0) \, \, .
\end{equation}
This law implies then a linear relation between $\Gamma$ and Log$E$ like
that apparent for the best fit lines plotted in Fig. \ref{fig3}, 
which have remarkably similar slopes corresponding to a curvature parameter 
$b$ between 0.13 and 0.16, the former corresponding to P2 and the latter to Ip. 

We are also interested in the spectral energy distribution, defined as $E^2F(E)$,
which for a log-parabolic law has a maximum at the energy:
\begin{equation}\label{eq3}
E_p= 10^{(2-a)/2b} \, \, .
\end{equation}

Mineo et al. (2006) used a log-parabola for the phase resolved spectroscopy
of the INTEGRAL observations of Crab and found a significant curvature which 
was found stable across the entire phase interval of the pulsed signal with
a mean value of 0.14$\pm$0.02.
We apply this law to each of the two components introduced by MCLM, as
described in detail in Sect. 6.2. 

\section{Shape and spectrum of the first peak}

Fig. \ref{fig4} shows the normalized profiles of P1 in the three energy 
bands 1.6--3 keV, 16--32 keV and 87--300 keV. 
As noticed by Mineo et al. (1997), there is an evident 
broadening which increases with energy.
In the lowest energy range the peak profile is asymmetric with a leading
side higher than the trailing one, as observed at optical frequencies 
(see, e.g., Kanbach et al. 2003).
This asymmetry changes in the hard X-rays as the trailing side becomes
progressively higher. 
The broadening, however, can be safely detected also on the leading side. 
We verified that this effect is also present to the same extent in the RXTE 
data and therefore it cannot be due to possible phase misalignements of 
the various BeppoSAX data sets.

\begin{figure}
\resizebox{\hsize}{!}{\includegraphics[height=8cm,width=8cm,angle=-90]
{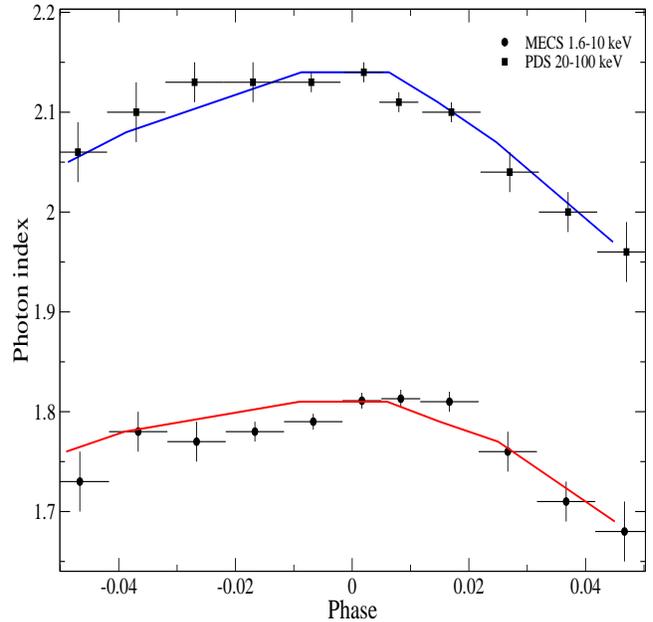}}
\caption{Fine phase resolved photon index across P1 in the 1.6--10 keV 
(MECS) and 20--100 keV (PDS) energy ranges. Solid lines are the photon 
indices derived using the two-component model as explained in Sect 6.4.
}
\label{fig5}
\end{figure}

Pravdo, Angelini \& Harding (1997), using a RXTE observation of the Crab 
pulsar, found that the central part of P1 has a spectrum softer than the 
wings.
In MCLM we confirmed this result from MECS data, but it was after 
questioned by Vivekanand (2002), who in a new analysis of the RXTE data 
in the range 5--60 keV obtained a stable value of $\Gamma$ across P1.
We performed a detailed analysis with a finer phase
resolution and confirmed the previous results: 
the photon index in the MECS range changes from $\sim$1.7 in the wings of 
P1 to 1.82 in the core, as shown in the plot of Fig. \ref{fig5}.
The same effect is also apparent in the PDS data.
As it will be clearer in the following section this result is relevant
to explain the total pulsed emission as due to the superposition of
two components with different spectra. 

\section{The two component model}

To explain at the same time the change of the pulse profile and the phase 
evolution of the photon index, in MCLM we proposed a simple model in which 
the observed pulsed emission is considered as the superposition of two 
components having different phase distributions and energy spectra.
The first component was assumed to have the same pulsed profile observed at 
optical frequencies (Smith et al. 1988), with P1 
much more prominent than P2 and a very low intensity in the Ip region.
The reason for this choice is that a similar shape is also observed at 
energies higher than 30 MeV, despite the profile of P2 being slightly 
different (Fierro et al. 1998).
This component was called `optical' (shortly $C_{O}$).
The other `X-ray' component $C_{X}$ reaches the greatest relative 
intensity in the hard X to low-energy $\gamma$-rays and it is necessary
to explain the change of the P2/P1 ratio. 
We derived its phase profile to reproduce the observed one when 
it is added to $C_{O}$: 
we found that $C_X$ increases monotonically up to the phase 0.4 and then 
it has a rather sharp cut-off (Fig. 6).
The main properties of the latter component were then estimated from the 
BeppoSAX data, with a fitting procedure of several 
pulse profiles at different energies, as explained in detail in MCLM. 

In this new analysis we followed the same approach and considered again
the same two components with only slight changes, as explained in 
the following subsections.

\begin{figure}
\resizebox{\hsize}{!}{\includegraphics[height=10cm,width=10cm]{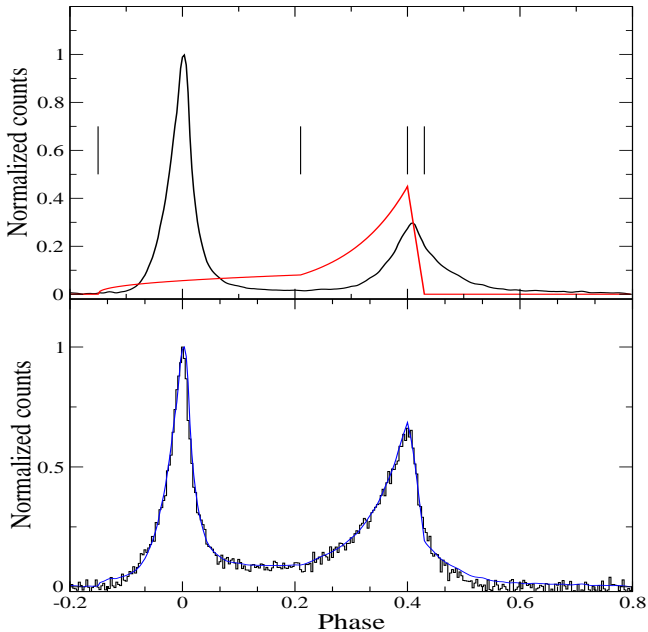}}
\caption{Upper panel: phase distributions of the components $C_{O}$ 
and $C_X$ with the proper normalizations. $C_O$ is derived from OPTIMA 
lightcurve (Kanbach et al. 2003). The four vertical bars indicate 
the phases from $f_1$ to $f_4$ used in the analytical model of $C_X$.
Lower panel: comparison between the observed pulse profile in the energy 
interval (8.0--10.0) keV (MECS data) and that obtained using the two 
component model computed for the mean energy of 8.85 keV. 
}
\label{fig6}
\end{figure}

\subsection{The components' shapes}

We obtained the phase profile of $C_O$ from the recent light curve 
observed with the high speed photo-polarimeter OPTIMA with a time 
binning of 112 $\mu$s (Kanbach et al. 2003), comparable to that used 
for the BeppoSAX profiles.
We used a high resolution digital scanner to convert the OPTIMA data
and then we applied a rebinning to have the final phase profile with
300 bins, the same used for the X-ray data. This profile is shown
in the upper panel of Fig. 6.
 
The pulse shape of $C_X$ can be obtained by the difference between the
observed profiles and $C_O$.
However, using the same approach as in MCLM we preferred to model $C_X$ 
by means of an analytical expression to make simpler computations.
$C_X$ was factorized in two functions:
\begin{equation}
F_X(E,f)=Y(E)~g(f,E) \,\,,
\end{equation}
where $Y(E)$ 
is assumed to depend only on the photon energy $E$ and represents 
the spectral distribution of the component to be derived from the data
as in MCLM, while $g(f,E)$ describes the phase behaviour and depends on 
both variables.
In MCLM we found that the assumption that $g$ is only a 
function of the phase was not fully satisfied and a moderate change
of the phase profile with the energy will be considered in this paper
for a more precise description of the data. 
The whole phase interval of $C_X$ is denoted by ($f_1$,$f_4$) 
and it is divided by two inner points at the phases $f_2$ and $f_3$ 
into three segments: in ($f_1,f_2$) the $C_X$ shape is given by
a power law, in ($f_2,f_3$) by an exponential function joining the 
first one at $f_2$, and finally a linearly descending branch from $f_3$ 
to $f_4$ that connects the maximum to the zero level of the off-pulse. 
We have then:

\begin{equation}
g(f,E)= \exp\{p(E) (f_2-f_3)\}\left(\frac{f-f_1}{f_2-f_1}\right)^{s}\,\,,
\end{equation}

\noindent
for $f_1<f<f_2$, and

\begin{equation}
g(f,E)=\exp\{p(E) (f-f_3) \}\,\,,
\end{equation}

\noindent
for $f_2<f<f_3$.

We emphasize that $C_X$ is non zero also in the phase interval of P1: 
this choice is necessary to explain the broadening of this 
peak with increasing energy and the softer spectrum of the peak centre 
with respect to both wings (see Sect. 5.1). 
Note, however, that because of the dominance of P1 with respect to Ip 
in $C_O$, is it difficult to establish the actual shape of $C_X$ across P1.
Our assumption that it has the smooth power law shape of Eq. (5) must be 
considered as one of the simplest approximations.   

In MCLM we found that the phase interval boundaries, equal to 
$f_1=-0.15$, $f_2=0.21$, $f_3=0.40$, $f_4=0.43$ and the power law 
exponent $s=0.4$ can be considered constant with the energy, whereas 
a better agreement with the data is obtained if $p$ is allowed to vary, 
changing from about 11 in the keV range to 8 above 150 keV.
A simple regression gives the relation:  
\begin{equation}
p(E) = -1.5811~\ln(\mathrm{Log E}) + 9.318
\end{equation}
that we used in our calculations.
An example of how these two components combine to reproduce
the observed profile in the energy range 8--10 keV is shown in Fig. 
\ref{fig6}.

Even though this model has some minor defects of accuracy, it gives an 
acceptable picture of the broad-band spectral and phase properties of 
the pulsed emission of Crab, as we will show in the following sections.
This model is not only a useful tool to 
represent a limited data set and it can be successfully used to 
extrapolate the pulse shape and spectra outside the energy range in which 
it is established.
Note that Kuiper et al. (2001) derived a seven-bin phase profile of their 
component necessary to explain the spectral difference between Ip and the 
peaks and found a structure similar to $C_{X}$. 

\begin{figure}
\resizebox{\hsize}{!}{\includegraphics[angle=-90]{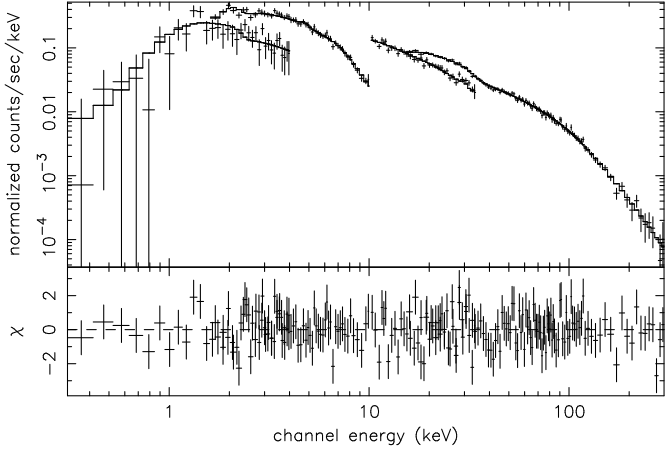}}
\caption{Spectral best fit of the four BeppoSAX-NFI data sets of 
Ip with the log-parabolic law of Eq.(9).}
\label{IP_lp}
\end{figure}

\subsection{Spectral properties of the two components}
To model the phase and spectral properties of the total emission as
the superposition of $C_O$ and $C_X$ we had to estimate their
spectral distributions.
The low intensity of $C_O$ in the Ip interval can be used to represent
its X-ray flux as due only to the $C_X$, 
while in the first and second peaks both components contribute to the
total emission, although with different relative intensities. 
Assuming that the spectrum of each component is given by a log-parabolic
law, we have for P1, Ip and P2:
\begin{equation}
F(E)_{P1} = (A_O)_{P1} E^{-(a_O+b_O\mathrm{Log}E)} +  
(A_X)_{P1} E^{-(a_X+b_X\mathrm{Log}E)}
\end{equation}
\begin{equation}
F(E)_{Ip} = (A_X)_{Ip} E^{-(a_X+b_X\mathrm{Log}E)}
\end{equation}
\begin{equation}
F(E)_{P2} = (A_O)_{P2} E^{-(a_O+b_O\mathrm{Log}E)} + 
(A_X)_{P2} E^{-(a_X+b_X\mathrm{Log}E)}
\end{equation}

The parameters to be determined are thus  $a_O, b_O, a_X$ and $b_X$, 
besides the normalization factors $(A_O)_{P1}, (A_O)_{P2}$ and $(A_X)_{Ip}$. 
The constants $(A_X)_{P1}$ and $(A_X)_{P2}$ can be obtained from $(A_X)_{Ip}$ 
and from the shape $g(f,E)$ of $C_X$.
Let $F_X(E,f)$ be the flux of $C_X$ component; we have:
\begin{equation}
\frac{\int_{Ip} {F_X(E,f)df}}{\int_{P1,P2} {F_X(E,f)df}} = 
\frac{(A_X)_{Ip}}{(A_X)_{P1,P2}} \, \, .
\end{equation}
With the factorization of Eq. (4), we obtain:
\begin{equation}
(A_X)_{P1, P2} = (A_X)_{Ip}\frac{\int_{P1,P2} {g(f,E)df}}{\int_{Ip} {g(f,E)df}} \, \, .
\end{equation}
Then, integrating $g(f,E)$ in the proper phase intervals of Ip, P1 and P2, and 
obtaining $(A_X)_{Ip}$ from a spectral fit, we can evaluate $(A_X)_{P1}$ 
and $(A_X)_{P2}$.

\begin{figure}
\resizebox{\hsize}{!}{\includegraphics[angle=-90]{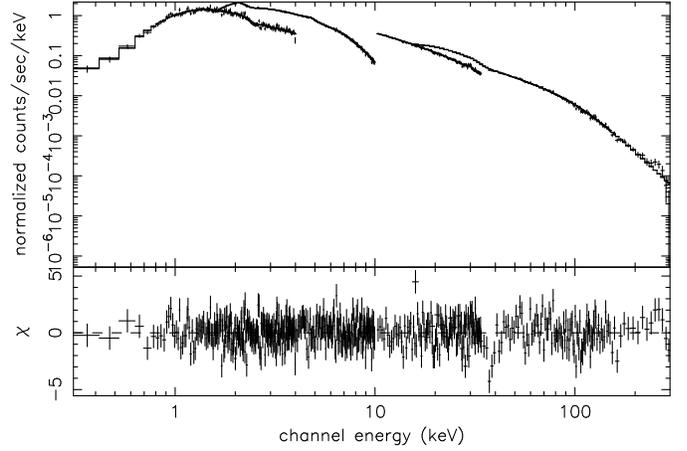}}
\caption{Spectral best fit of the four BeppoSAX-NFI data sets of 
P1 with the combined log-parabolic model of Eq.(8) and with the UV 
constraint of Eq.(12).}
\label{P1_lp}
\end{figure}

\begin{figure}
\resizebox{\hsize}{!}{\includegraphics[angle=-90]{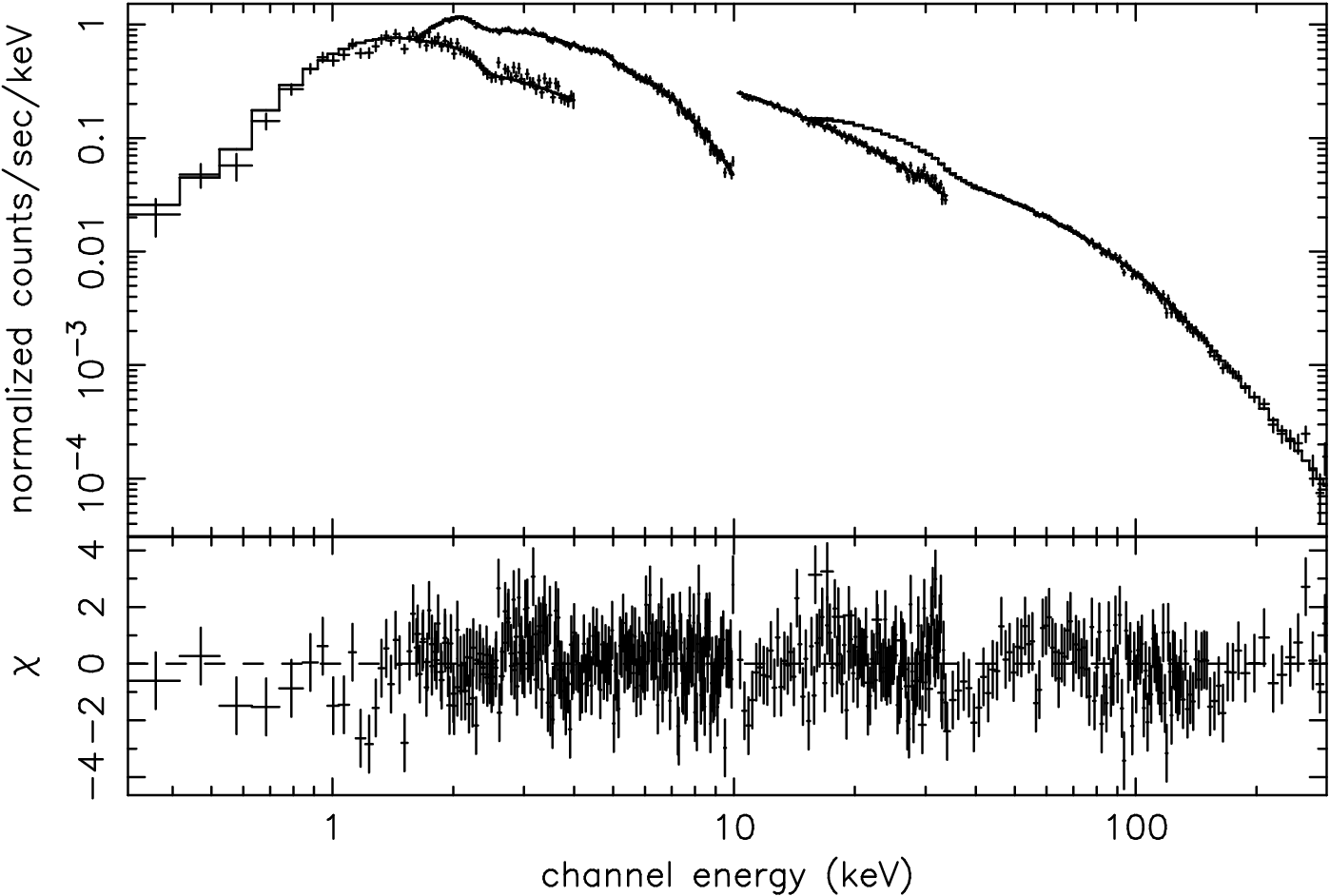}}
\caption{Spectral best fit of the four BeppoSAX-NFI data sets of 
P2 with the combined log-parabolic model of Eq.(10) and with the UV 
constraint of Eq.(13).}
\label{P2_lp}
\end{figure}

\begin{table*}
\caption{ Parameters of the two-component, log-parabolic spectral model fits 
of BeppoSAX data (LECS, 0.1--4.0 keV; MECS 1.6--10.0 keV; HPGSPC 10--34 keV; 
PDS 15--300 keV).}
\label{tab:logparfits}
\begin{center}
\begin{tabular}{cccc}
\hline
Parameter & First peak (P1) & Interpeak (Ip) & Second peak (P2) \\
\hline
$a_O$ & 1.652 $\pm$ 0.003 & --- & 1.670 $\pm$ 0.008           \\
$b_O$ & 0.160 $\pm$ 0.001 & --- & 0.164 $\pm$ 0.001          \\
$a_X$ & --- & 1.28 $\pm$ 0.04  & ---            \\
$b_X$ & ---  & 0.16 $\pm$ 0.01 & ---          \\
$A_O$($^*$) & 0.1660 $\pm$ 0.0005 & --- & $(4.72 \pm 0.06)\cdot 10^{-2}$     \\
$A_X$($^*$) & --- & $(2.3 \pm 0.1)\cdot 10^{-2}$ & ---            \\
$\chi^2_r$ & 1.02 (867 d.o.f) & 0.99 (866 d.o.f) & 1.09 (867 d.o.f)         \\
\hline
\\
$(^*)$~ photons cm$^{-2}$ s$^{-1}$ keV$^{-1}$
\end{tabular}
\end{center}
\end{table*}

To obtain a fully consistent picture, in addition to the X-ray data, 
the spectral distribution of $C_O$ must satisfy the optical-UV observations.  
Sollerman et al. (2000), from STIS/HST observations, found that the UV 
spectrum of the Crab pulsar is remarkably flat for both peaks and can be
described by a single power law, with an energy index $\alpha_{UV} = -0.11$.
To better constrain our spectral parameters we impose the condition that
the extrapolation of the $C_O$ log-parabolic spectral distribution in the 
UV range would be consistent with this value and from Eq. (2) we derived
the following relation between $a_O$ and $b_O$:
\begin{equation}
b_O = 0.21~a_O - 0.187
\end{equation}
which was included in the spectral best fitting procedure.
The resulting best fit spectra with the model of Eqs. (8)-(10) and the 
residuals for Ip, P1 and P2 of all NFI data sets of BeppoSAX are
shown in Figs. \ref{IP_lp}, \ref{P1_lp}, \ref{P2_lp}, respectively,
and the best fit values are reported in Table  \ref{tab:logparfits}.

The condition of Eq. (13) should not be considered very stringent
in the evaluation of the spectral parameters.
In fact, when it is released and only X-ray data are used, the best fit
values of $a$ and $b$ are consistent with the ones in in Table 
\ref{tab:logparfits}.
Eq.(13), therefore, must be considered as a useful tool to reduce
the parameter space and to avoid possible inconsistencies between the
optical-UV and X-ray data.
Moreover, the actual spectral distribution at frequencies much lower
than that of the peak can deviate from a pure log-parabola and can 
approximate a less curved law like a simple power law.
Unfortunately, in this frequency range extinction effects are important
(see Sect. 3.2) and it is not simple to reach a satisfactory picture.

Three other Crab-like pulsars show curved 
X-ray spectra that can be fitted by log-parabolae with curvature
parameters very similar to that of Crab: PSR B1509$-$58 ($b$ = 0.16$\pm$0.04, 
Cusumano et al. 2001), PSR B0540$-$69 ($b$=0.143$\pm$0.003, de Plaa,
Kuiper \& Hermsen 2003) and PSR J0537$-$6910 ($b$=0.15$\pm$0.05, Mineo,
Cusumano \& Massaro 2004). 
This finding can be considered an indication that the same radiation 
mechanism is at work in these sources. 

We recall that there are important differences between our modelling 
and the one of Kuiper et al. (2001). 
These authors assumed that the pulsed signal observed above $\sim$30 MeV,
modeled by a single power law, extrapolates down to the X-ray range and 
therefore fitted the phase resolved spectra by means of this law plus
two log-parabolae. 
The estimate of the curvature parameters is thus affected by the power law, 
and their values (0.299 and 0.084) are quite different from those found
by us and cannot be directly related to the slopes of the linear best fits
of Fig. 3.
The normalization factors of these components are established independently
in each phase interval considered, increasing the number of free parameters,
whereas we evaluated the normalization factors of $C_O$ and $C_X$ from 
their phase distributions, as explained above. 

\subsection{The X-ray shape and the spectral index of P1}

The two componenent model is founded only on data from the optical to
the hard X-ray range and, in principle, it could fail in describing some
specific pulsar properties and the spectral behaviour outside this energy 
range. 
A useful test is to verify how the model reproduces the characteristics 
of P1 described in Sect. 5.
Fig. 4 shows the model profiles computed at the three energies of 2.0, 22 
and  85 keV, approximately the mean values of the considered energy intervals.
The pulse broadening is well described and the only deviation is between the 
phases 0.025 and 0.075, where the data show a small excess, which is
s likely due to the simple analytical shape adopted for $C_X$. 
The search for a better agreement would likely require a more complex formula 
with a greater number of parameters without a clear improvement of the general 
description of the pulsed emission. 

A simple approximate way to evaluate the mean photon index from the 
model is to compute the ratio of the summed intensities of 
$C_O$ and $C_X$ at the two extremes of the energy range:
\begin{equation}
\Gamma_{f}= - \frac{Log(F(E_2)/F(E_1))}{Log(E_2/E_1)} + \Gamma_O \, \, .
\end{equation}
Since we used normalised pulse profiles, this spectral index would be
 zero at the centre of P1 and therefore we added the constants 
$\Gamma_O$=1.83 and 2.14 for the MECS and PDS data, respectively, to 
match the measured values. 
In this case the agreement between the observed points and the model, 
shown in Fig. 5, is generally satisfactory, in particular the asymmetric variation
of $\Gamma(f)$ with the softest value at the P1 centre. 
Both these results are due to the $C_X$ contribution in the phase interval of P1 
necessary to represent the pedestal apparent in the hard-X to low-energy $\gamma$ rays.
These results show that observations are well reproduced if the spectrum of 
this pedestal is similar to that of Ip, and therefore it is reasonable to consider 
them all as belonging to the same component.

\subsection{The total pulse profile in the MeV range}

Another important verification of the model is the comparison
of pulse profiles and spectra at energies higher than the BeppoSAX-NFI 
range.
In particular, we considered the low-energy $\gamma$-ray data obtained by 
IBIS-PICsIT on board INTEGRAL (Kuiper et al. 2003) and by COMPTEL-CGRO 
(Kuiper et al. 2001).

We computed the expected pulse shapes from the two component model 
at the three energies 0.306, 0.830 and 1.90 MeV, which are inside the 
energy ranges of the considered pulse profiles (0.260--0.364), (0.75--1.0)
and (1.0--3.0) MeV.
We used the spectral parameters of Table 4 and the value of $p$ derived 
from Eq. (7).
The resulting phase histograms are shown in the three panels of Fig. 10
together with the data.
We see that the model profiles match well the most important 
features such as the heights of the two main peaks and gives an acceptable 
representation of the Ip region. 
Differences are generally within the observational uncertainties, because of 
the statistics available in these bands. 
We see therefore that our model is able to extrapolate well the pulsed emission 
up to the MeV range.

\subsection{The emission at optical-UV frequencies}
For a fully consistent description of the Crab pulsed emission we evaluated 
how the spectral laws derived in the previous Sections extrapolate in the 
optical-UV range.
At these frequencies only the contribution expected from $C_O$ is important, 
that of $C_X$ being smaller than one order of magnitude.
A simple extrapolation of the model gives a flux lower than that derived
by the optical-UV observations by Percival et al. (1993) and Sollerman et al. 
(2000).
As explained in Sect. 3.2, however, reddening effects are very important and
even a slight change of the extinction parameter can modify the resulting 
spectrum.

In order to match the optical points it is not possible to reduce the value 
of $b$ because it  would be inconsistent with the X-ray data.
The simplest way is to assume that at frequencies lower than $\sim$0.1 keV
the log-parabola tends to approach a power law.
A possible spectral distribution, including optical-UV points, is shown in 
Fig. 11.

\begin{figure}
\resizebox{\hsize}{!}{\includegraphics[angle=0]{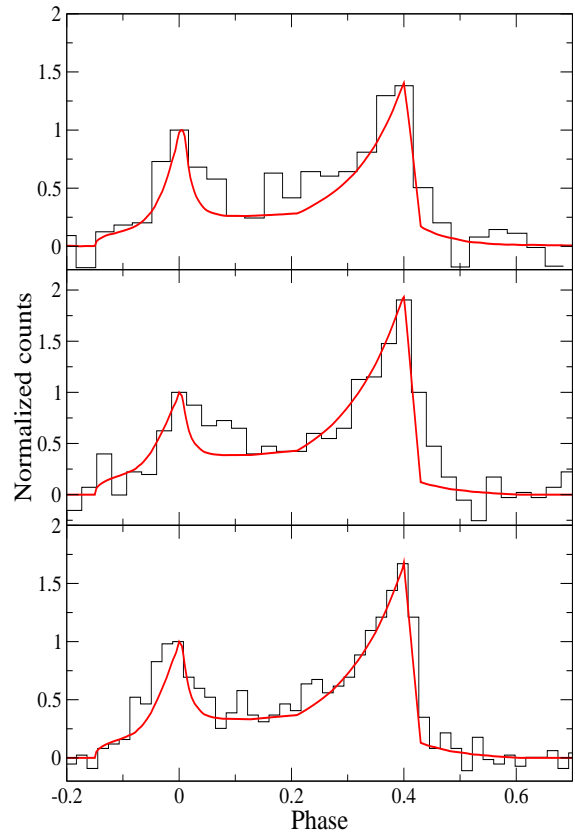}}
\caption{Comparison between the Crab pulse profiles in the energy
ranges 0.260--0.364 (IBIS-PICsIT data from Kuiper et al. 2003,
upper panel), 0.75--1.0 MeV and 1.0--3.0 MeV (central and lower panel, 
COMPTEL-CGRO data from Kuiper et al. 2001) computed using the two 
component model.}
\label{comptel_mod}
\end{figure}

\section{Extension of the two-component model at higher energies: 
the need for other components}

At energies higher than $\sim$30 MeV the model developed in the previous 
section based only on the two components $C_O$ and $C_X$ is not 
adequate to describe the emission properties of the Crab pulsar.
It has been known since early observations that the $\gamma$-ray pulse shape 
is similar to that of $C_O$, although some minor differences 
are present.
The results of COMPTEL and EGRET observations (Kuiper et al. 2001) 
provided high energy pulse profiles and spectra good enough to develop 
a more complete model to describe the overall properties of Crab. 
We recall that there are some similarities between the properties of
the pulsed emission in the hard X-rays and in the high energy 
$\gamma$-rays.
Kanbach (1999) showed that the relative flux of P2 to that of P1 
increases with energy.
The pulse profile above 1 GeV compared to that above 0.5 GeV
shows a clear excess of Ip and P2 with respect to P1 (see, e.g., Fig. 
12 of Kuiper et al. 2001), 
Moreover, above $\sim$5 GeV, P2 seems to be the only dominant
feature in the pulse profile, despite the small number of detected 
events (Thompson 2004).

To explain these properties in a consistent way, from the optical 
to the GeV band, 
we extend the previous two component model assuming 
that there are two other high-energy spectral components, hereafter 
indicated by $C_{O\gamma}$ and $C_{X\gamma}$, which are strictly 
related to $C_O$ and $C_X$, respectively, because their pulse shapes 
are similar and the spectral distributions are approximated by 
log-parabolic laws, but shifted in energy.
These additional two components imply at least another six adjustable 
parameters, i.e. the peak energies, curvatures and normalizations of the two 
log-parabolae, but the statistical quality of the available data does
not allow a good estimate of all of them.
We assumed, therefore, that their curvatures are equal to that of the 
corresponding low energy components ($b=0.16$) and found peak 
energies and normalization factors able to describe well the observed 
spectra.

\begin{figure}
\resizebox{\hsize}{!}{\includegraphics[angle=-90]{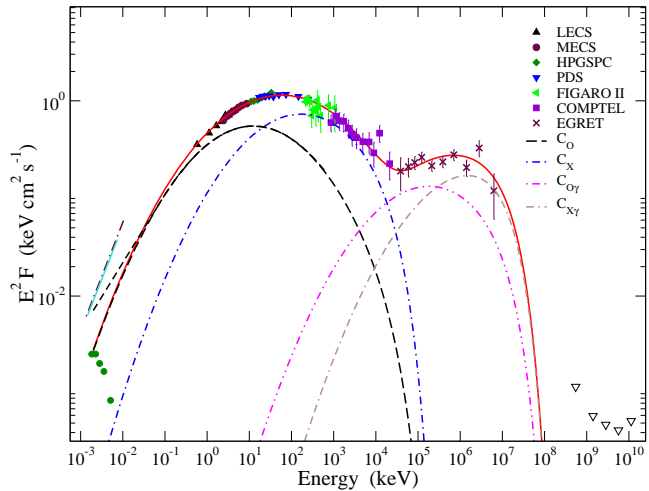}}
\caption{The broad-band spectral energy distribution of the total averaged 
pulse of the Crab Pulsar, with the four components of the model. 
Data points are from BeppoSAX, Compton-CGRO and FIGARO II.
We also included TeV upper limits (downward open triangles) from Aharonian et al. 
(2004) and the optical data (green circles) from Percival et al. (1993), together with the 
best fit after reddening corrections (solid cyan line). Also shown (brown dash-dotted line) is the optical dereddened data from Sollerman et al. (2000).
A possible low frequency extrapolation of the two component SED to match these
optical data is plotted.
}
\label{modello_plot_TOT}
\end{figure}

We first adapted this four component model to the total pulsed spectrum 
and evaluated the best normalization factors of $C_{O\gamma}$ and 
$C_{X\gamma}$.
First we noticed that to be consistent with the upper limits to the pulsed 
emission in the TeV range (e.g. Lessard et al. 2000, Aharonian et al. 
2004) it was necessary either to increase the values of $b$ or to
include an exponential cut-off in the $C_{X\gamma}$ spectrum. 
We preferred to follow the latter approach, because there is no simple
way to fix the curvatures, and found that a cut-off energy $E_c = 15$ 
GeV gives a reasonable fit, although this value cannot be precisely 
evaluated.
The same cut-off energy was also assumed for $C_{O\gamma}$, although
it was not required by the data, because the spectral steepening 
of the log-parabola reduces the relevance of this component above 
a few GeV.
The resulting spectral energy distribution compared with BeppoSAX,
COMPTEL and EGRET data is shown in Fig. \ref{modello_plot_TOT}.
We also added some data points obtained from the FIGARO II experiment
(Agrinier et al. 1990, Massaro et al. 1998) to fill the gap between 300 
keV and 1 MeV.

\begin{figure}
\resizebox{\hsize}{!}{\includegraphics[angle=-90]{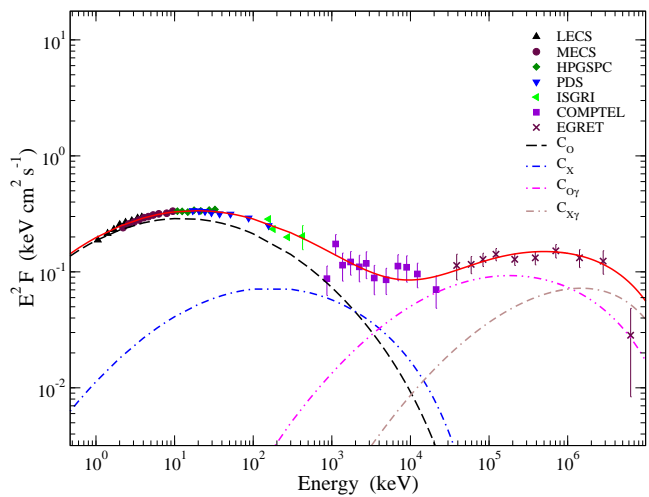}}
\caption{The broad-band spectral energy distribution of P1 with the four 
components of the model. 
Data points are from BeppoSAX, INTEGRAL-ISGRI and Compton-CGRO.}
\label{modello_plot_P1}
\end{figure}

\begin{figure}
\resizebox{\hsize}{!}{\includegraphics[angle=-90]{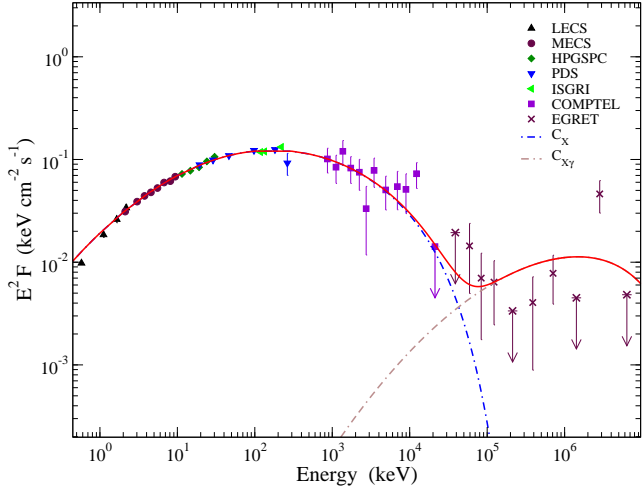}}
\caption{The broad-band spectral energy distribution of Ip with the 
components of the model. 
Data points are from BeppoSAX, INTEGRAL-ISGRI and Compton-CGRO.}
\label{modello_plot_IP}
\end{figure}

\begin{figure}
\resizebox{\hsize}{!}{\includegraphics[angle=-90]{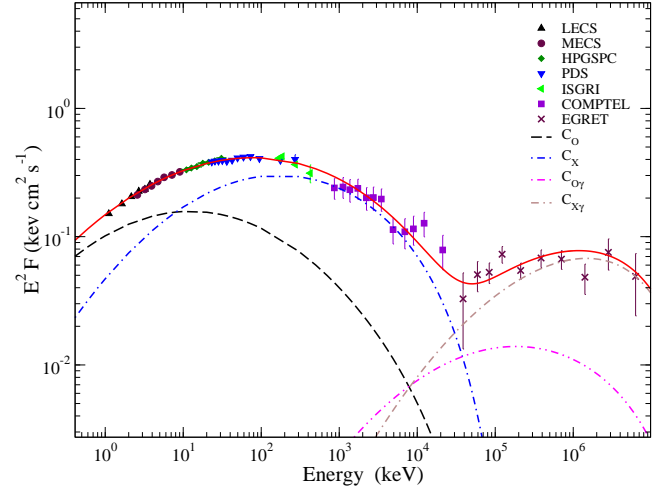}}
\caption{The broad-band spectral energy distribution of P2 with the four 
components of the model. 
Data points are from BeppoSAX, INTEGRAL-ISGRI and Compton-CGRO.}
\label{modello_plot_P2}
\end{figure}

The spectra of P1, Ip, and P2 were computed using the phase intervals
of Kuiper et al. (2001) and the intensity ratios derived from the pulse 
shapes of $C_O$ and $C_X$. 
Figs. \ref{modello_plot_P1}, \ref{modello_plot_IP} and \ref{modello_plot_P2} 
show the resulting spectral energy distribution together with the data derived 
from the literature and our new analysis of BeppoSAX and ISGRI-INTEGRAL data.
The agreement with the data in the P1 and P2 ranges is satisfactory, 
confirming that in these intervals the pulse shapes of the $C_{O\gamma}$ and 
$C_{X\gamma}$ components are similar to those of $C_{O}$ and $C_{X}$, 
respectively, consistent with our assumption. 
The poor statistics of data in the Ip region does not allow a good enough
verification of the model.
The SED of Ip has only two components and is poorly known above 10 MeV.
However, the model expectation for the Ip intensity was significantly 
higher than the observational points.
Consequently, we reduced the normalization of $C_{X\gamma}$ by a factor 
of about 2 to approximately match the data. 
This indicates that the profile of $C_{X\gamma}$ in this range must be
different from $C_{X}$, and it does seem to be sharper.
When the same change is applied to the spectrum of P1 we did not find
a significant modification, the relative contribution of $C_{X\gamma}$ being
much lower than $C_{O\gamma}$.
Note in the SEDs of P1 and P2 how the mild spectral curvature is well 
reproduced by the superposition of the four components and  
that the positions of the minima are not the same: it is around 
10 MeV for P1 and around 40 MeV for P2. 
At present, however, the data do not allow an accurate estimate of this 
difference.
New observations with richer statistics will be useful to constrain 
 this feature more precisely, to provide further support for our 
multi-component model and to evaluate the component parameters.
All the main parameters of the full model are given in 
Table \ref{tab:mcmodel}.

\begin{table}
\caption{Main parameters of the four component model of the pulsed emission
of the Crab pulsar.}
\label{tab:mcmodel}
\begin{tabular}{cccc}
\hline
Component & $E_p$ & $E_c$ & $L_{bol}(^*)$    \\
          & MeV   & GeV & erg/s  $10^{36}$ \\
\hline
$C_O$     & 12.2$\cdot 10^{-3}$ & --- & 2.8   \\
$C_X$ & 178$\cdot 10^{-3}$ & ---  & 3.8 \\
$C_{O\gamma}$ &  300 & $\leq$15  & 0.97 \\
$C_{X\gamma}$ & 2000 & $\sim$15 & 0.92  \\
\hline 
\end{tabular}

$(^*)$~ assuming isotropic emission ($\omega_a = 4\pi$). 
\end{table}

This model is not univocally determined by data. 
It is apparent from Fig. 12 that in COMPTEL and EGRET data the P1 
spectrum can be fitted by a single power law with the only exception being
the point at the highest energy where a cut-off may be present, as already 
proposed by Kuiper et al. (2001).
In the spectral plots evaluated by these authors assuming a power law 
component with a phase independent spectral index, one can see that EGRET
data for P1 are all in excess with respect to the best fit, while practically
all those of COMPTEL lie under it.
Moreover, large systematic deviations from this best fit are apparent
in the leading and trailing wings of P1 and P2 (see also Fierro et al. 1998).
We conclude that the model based only on components having spectral distributions
approximated by a log-parabolic law gives a simpler picture of this complex
spectral behaviour.

\begin{figure}
\resizebox{\hsize}{!}{\includegraphics[angle=-90]{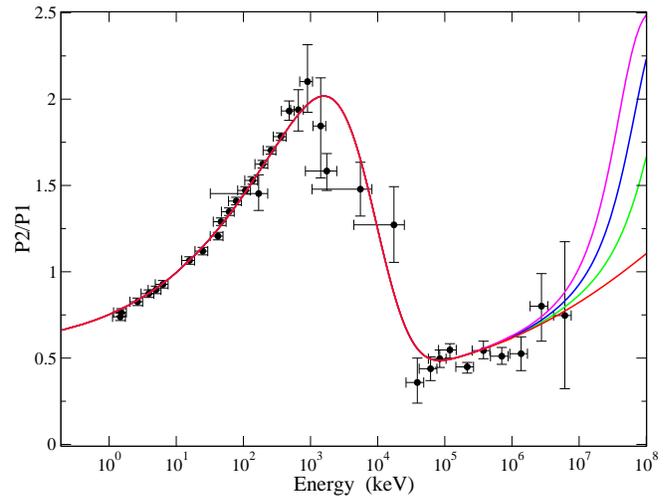}}
\caption{The ratio between the fluxes of P2 and P1 phase regions (P1: 
-0.06--0.04; P2: 0.32--0.43), compared to the predictions of the model. 
The data points come from various experiments (Kuiper et al. 2001).
The various extrapolations above 1 GeV correspond to different values 
of the cut-off energy of the $C_{O\gamma}$ spectrum: 15 GeV (red), 13 GeV (green),
11 GeV (blue) and 9 GeV (violet) - from bottom to top.}
\label{rappP2P1}
\end{figure}

\begin{figure}
\resizebox{\hsize}{!}{\includegraphics[angle=-90]{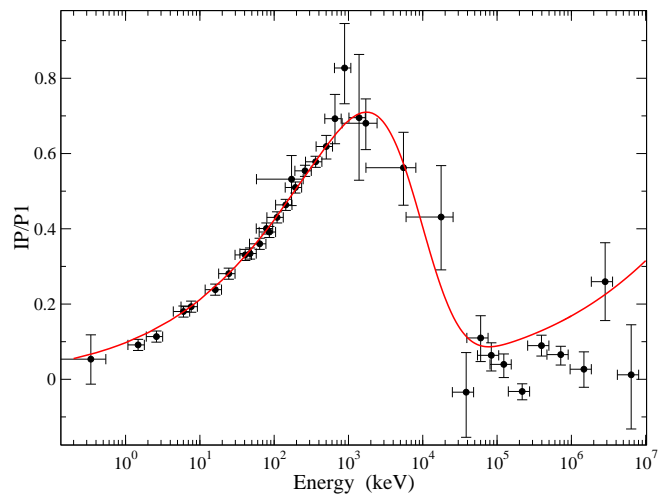}}
\caption{The ratio between the fluxes of Ip and P1 phase regions (P1: 
-0.06--0.04; Ip: 0.14--0.25), compared to the predictions of the model. 
The data points come from various experiments (Kuiper et al. 2001).}
\label{rappIPP1}
\end{figure}

Another test of the model is to verify how it reproduces the flux ratios
P2/P1 and Ip/P1, frequently considered in the literature to describe the 
pulsed emission of Crab.
We computed these ratios for the phase intervals adopted by Kuiper et al. 
(2001), and the results are plotted in Figs. \ref{rappP2P1} and \ref{rappIPP1} 
for P2/P1 and Ip/P1, respectively.
The general behaviours of both ratios are well reproduced, in particular 
the sharp decreases between 1 and 10 MeV and the following slow increase
to the GeV range. 
We stress that the sharpness of the decrease depends on the curvature and peak 
energies of log-parabolic spectra and of the adopted cut-off.

We also calculated the bolometric flux of each component by integrating 
the log-parabolic spectral distribution over all energies.
This integral can be easily computed analytically (Massaro et al. 2004) 
and the result is:
\begin{equation}
F_{bol} = \int_0^\infty A E^{-(a+b\mathrm{Log}E)}~dE = 
\frac{2.70}{\sqrt{b}}E^2_{p}F(E_{p}) \, \ .
\end{equation}
The luminosity of each of the four components is 
$L = F_{bol} \times d^2 \omega_a$, where $d$ is the distance to the Crab 
and $\omega_a$ is the solid angle occupied by the emission beam. 
We take $d\simeq2$~kpc $=6.2 \times 10^{21}$ cm, whereas the value of 
$\omega_a \leq 4\pi$ is difficult to estimate because it depends on 
the emission geometry, which is not completely known.
The resulting values for the isotropic case ($\omega_a = 4\pi$) are 
given in Table \ref{tab:mcmodel}, but they are upper limits to the  
actual values. 
The total bolometric luminosity, obtained by adding the contributions of the 
four components, is $L = 8.5 \times 10^{36}$ erg s$^{-1}$, which drops to
$\sim6.8\times 10^{35}$ erg s$^{-1}$ for $\omega_a = 1$ str. 

\section{Discussion}

The development of a detailed physical model for the spectral and phase 
distributions of the broad-band emission from the Crab pulsar is a 
difficult problem.
It requires a precise geometrical definition of the regions inside 
the magnetosphere where the observed radiation originates, and the knowledge of 
parameters like the orientation angles between the magnetic axis and the line of 
sight to the spin direction.
Several models have appeared in the literature based on either polar cap 
or outer gap geometries.
Usually, these models are focused on reproducing either the total 
spectrum or the phase profile and generally they are not fully satisfactory 
in explaining the complex observational picture.
Moreover, the possibility that the observed features of the pulsed signal 
can arise from the superposition of two or more distinct components is not
taken into account.

We followed another approach and searched for a possible interpretation 
of the Crab signal based on the superposition of two or more components 
that provides a consistent description of the spectral and phase 
distributions. 
MCLM showed that a possible explanation of the energy dependence
of the pulse shape of Crab in the soft to hard X-rays is that we 
are observing two emission components with different phase and 
energy distributions. 
In that paper we introduced an empiric model, based on a collection of 
BeppoSAX observations, covering the energy range from 0.1 to about 300 
keV, from which it was possible to estimate some properties of the main 
components.
Moreover, we showed that the X-ray spectrum of P1 presents a mild
curvature well fitted by a log-parabola.
At energies higher than 30 MeV, however, this two component model fails 
to represent both pulse profiles and spectra, as observed by EGRET-CGRO 
(Fierro et al. 1998).   
To take into account $\gamma$-ray data, Kuiper et al. (2001) proposed 
the existence of three components, two of them having log-parabolic spectra,
while the third one with a power law spectrum and a phase modulated intensity  
which reaches the highest level in correspondence of the two peaks and 
is almost absent in the Ip region.
According to our point of view, however, this hypothesis is not fully
consistent with the data, because the flux increases of Ip and P2 are 
very similar, suggesting that their X-ray emission is dominated by the
same physical mechanism. 
Moreover, the parameters describing the curvatures of their two log-parabolic 
distributions are very different, while the analysis of X-ray spectra reported in 
Sect. 6.2 indicates that the curvature is rather stable with phase at an
intermediate value between them. 
Finally, the power-law component does not match well the data in some
phase intervals at $\gamma$-ray energies where its contribution is dominant.

To achieve a more consistent scenario, in the present paper we propose
a model of the pulsed emission from Crab based on four components.
It is properly a two double-component model because each component 
pair has similar phase distributions and spectra shifted in energy.
At present, our model gives an empiric description 
of the broad-band properties and its validity is based on the limited number 
of assumptions and on the resulting capability to obtain a consistent 
description of the observations outside the energy ranges used to evaluate 
the parameters.
For instance, the model extrapolates well the pulse profiles in the MeV 
range (see Sect.6.4) and the change with energy of the P2/P1 and Ip/P1 ratios. 
We stress that Crab is not the only pulsar that shows a behaviour
that can be interpreted by a multicomponent model.
Harding et al. (2002) proposed that the X-ray pulsed emission from Vela 
originates from two non-thermal components, one coincindent in phase
with the $\gamma$-ray pulse profile and the other one with the optical.  

The physical processes at the origin of these components, and the location 
in the magnetosphere where they occur, must be further investigated.
The development of a detailed physical model of the high energy emission
in the Crab magnetosphere is beyond the aim of the present paper, however 
some general indications on it can be derived from our conclusions.  
The same phase distribution assumed for each pair of components, such as $C_O$ 
and $C_{O\gamma}$, suggests that their angular pattern and emission sites 
must be coincident or very close, otherwise the aberration effects would 
modify the pulse shapes.
However, this is not necessarily true because on the trailing last
open field line, aberration and propagation time effects would cancel
to form a caustic, as shown by Dyks \& Rudak (2003).

According to a widely accepted scenario, primary electrons accelerated in a 
magnetospheric gap emit high energy photons which produce $e^{\pm}$ pairs 
against the magnetic field.
These secondary particles emit synchrotron radiation in the optical 
to MeV energy range. 
A first possibility to be considered is that $C_O$ and $C_X$ are synchrotron 
radiation from two different places whereas the corresponding high energy 
components are inverse Compton upscattered photons.
A very early inverse Compton model for Crab was proposed by Zheleznyakov 
\& Shaposhnikov (1972), when the $\gamma$-ray emission properties were 
known very poorly, and another model was applied to the Vela pulsar by
Morini (1983). A more recent development is that of Cheng and Wei (1995). 
Their model is based on the outer gap geometry and assumes that optical to
hard X-ray photons are emitted by secondary $e^{\pm}$ pairs created
outside the accelerating region by high-energy primary curvature photons.
$\gamma$-rays are then emitted via a synchrotron self-Compton (SSC) process. 
In this case we expect that Compton scattering of hard X-ray photons occurs
mainly in the Klein-Nishina regime, and therefore the observed high energy 
cut-off gives an estimate of the maximum energy of electrons.
The fact that it is in the GeV range could be consistent with their origin
from magnetic pair production. 

There is, however, another possibility for the origin of the high energy
components. 
In a polar cap scenario these photons could be those
emitted by primary electrons via curvature radiation and not absorbed by
magnetic pair production throughout the magnetosphere.
The attenuation length for this process  can be approximated by
(Erber 1966): 
\begin{equation}
x(B_{\perp}, E_{\gamma}) = 2.3 \times 10^{-8} \frac{B_{cr}}{B_{\perp}}~ 
\exp\left[\frac{4 B_{cr}} {3 B_{\perp}} \frac{2 m c^2}{E_{\gamma}} \right]
\mbox{\ \ \ \ cm\,,}
\end{equation}
where $B_{\perp}$ is the transverse magnetic field seen by a photon of energy 
$E_{\gamma}$, and $B_{cr}$=4.414$\times$10$^{13}$ is the quantum critical field.
It is easy to verify that a mean free path of the order of 10$^4$ cm for 5 
GeV photons is obtained in a transverse field $B_{\perp}\simeq$ 7.5$\times$10$^8$ G
and that it increases very rapidly even for decreasing $E_{\gamma}$.
This cut-off in the curvature $\gamma$-ray spectra at GeV energies has been 
verified since early numerical calculations, including an accurate evaluation
of the absorption coefficient, for a polar cap acceleration  
(Salvati \& Massaro 1978, Massaro \& Salvati 1979).
 
Although our multi-component model gives a consistent picture of the pulsed 
emission from Crab it is not yet completely determined by observational data.
An important test will be the study of the spectra and pulse profiles
at energies higher than a few GeV: in particular, we expect that above
about 5 GeV, P2 would be the dominant feature, while a good measure of
the flux in the Ip region will be very useful to draw the phase structure
of $C_{X\gamma}$.  
High quality data in this energy range will be obtained with the LAT
telescope on board the GLAST mission to be operative next year.
Another interesting test of the model could be obtained from phase
resolved polarization measures in the X-ray range. 
We know that the optical linear polarisation of the Ip differs from 
those of P1 and P2 both in strength and direction 
(Smith et al. 1988; Kanbach et al. 2003).
According to our model the polarization of the P2 X-ray emission must
be more similar to that of Ip, because of the higher contribution of $C_X$. 
New generation high sensitivity polarimetry for X-ray astronomy, such as
that proposed by Costa et al. (2001), could be very useful.  
Finally, the observation of other young spin powered pulsars, if their high
energy emission is similar to Crab, could also be very useful because we expect 
to observe them from different directions and therefore to see other pulse shapes, 
depending on the various combinations of the two components.

\begin{acknowledgements}
We are grateful to M. Salvati for interesting comments and to L. Kuiper 
who kindly gave us COMPTEL and EGRET data.
This work has been partially supported by Universit\`a di Roma ``La~Sapienza''.
\end{acknowledgements}

\end{document}